\documentclass{wqcd03}                 
\usepackage{txfonts}

\confname{QCD@Work 2003 - International Workshop on QCD, Conversano, Italy, 
14--18 June 2003}

\title{Observation of two new narrow mesons in the $D^{+}_{s}\pi^{0}$ and $D^{+}_{s}\pi^{0}\gamma$ final states. \\ Results from BaBar, Belle and CLEO.}

\author{Alexis~Pompili\thanks{Talk at the International Workshop on QCD: QCD@Work 2003 - Conversano (Italy) 14-18 June 2003 (eConf C030614).} }

\address{Dipartimento Interateneo di Fisica and INFN, Bari, Italy}

\newcommand{\gevcc}{GeV$/c^{2}$ }
\newcommand{\gevccc}{GeV$/c^{2}$}
\newcommand{\mevcc}{MeV$/c^{2}$ }
\newcommand{\gevc}{GeV$/c$ }

\newcommand{\mevccc}{MeV$/c$}

\begin{document}

\begin{abstract} The BaBar experiment has discovered a narrow state, denoted as $D^{*}_{sJ}(2317)^{+}$, near $2.32$ \gevcc in the $D^{+}_{s}\pi^{0}$ invariant mass distribution from inclusive $e^{+}e^{-}$ interactions at center-of-mass energies near $10.6$ GeV. The same experiment has hinted the presence of a second new state near $2.46$ \gevcc in the $D^{+}_{s}\pi^{0}\gamma$ mass spectrum. These discoveries have triggered CLEO and Belle experiments in their own search for new states coupled to $D^{+}_{s}$ meson. 
They have confirmed the existence of the $D^{*}_{sJ}(2317)^{+}$ state and established the evidence for the second state denoted as $D^{*}_{sJ}(2458)^{+}$. Belle has reported evidence for both states not only in continuum $e^{+}e^{-}$ annihilations but also in $B$ decays. These two new narrow resonances are characterized by masses significantly lower than predicted by potential models and enough to close off the most natural decay modes, thus forcing isospin-violating transitions. Their discovery has been a surprise because it contradicts the expectations of apparently established theory. This circumstance has led to a great effort on theoretical side as well.
\end{abstract}

\maketitle

\section{Introduction}
Potential models of mesons made up by an heavy quark and a light one have had so far reasonable success in describing the spectroscopy of $D$, $D_{s}$, $B$ and $B_{s}$ systems~\cite{potmod}. The spectroscopy of $c\overline{s}$ states is straightforward in the limit of large charm quark mass~\cite{csbar}: the total angular momentum of the light quark, $\vec{j} \! = \! \vec{l} + \vec{s}$, obtained by summing its orbital and spin angular momenta, is conserved. The $P$-wave states, all of which have positive parity, then have $j \! = \!3/2$ or $j\! = \! 1/2$; after combination with the heavy quark spin ($\vec{J} \! = \! \vec{j} + \vec{s_{h}}$), the former gives total angular momentum $J\! = \! 2,1$ whereas the latter gives $J \! = \! 1,0$. 
The $2^{+}$ and one $1^{+}$ state, forming the doublet associated to $j \! = \! 3/2$, are expected to have a small width because their dominant (OZI- and isospin-favoured) decays to $D^{*}K$ and $DK$ respectively would proceed via $D$-wave. On the contrary, the $0^{+}$ and the other $1^{+}$ state, associated to  $j \! = \! 1/2$, are expected by most potential models to have comparable masses (in the range $2.4 \! \div \! 2.6$ \gevccc) and to decay by kaon emission to the same final states, but with large widths (at least $200 \div \! 300$ MeV) since these decays would proceed via $S$-wave. 

The experimental information on the $S$-wave ground state is well established. The $D^{*+}_{s}$ meson is identified with the excited $1^{-}$ triplet state at $2112$ \mevcc, decaying to the $D^{+}_{s}$ meson identified with the ground $0^{-}$ singlet state at $1969$ \mevcc, mainly by photon emission. An isospin-violating pion emission also occurs with a rate of about $6\%$. The experimental information concerning the $P$-wave states is limited instead. The observations of the $D^{*}_{sJ}(2573)^{+}$ (with $J^{P}$ consistent with $2^{+}$) and the $J^{P} \! = \! 1^{+}$ $D_{s1}(2536)^{+}$ identified with the $j=3/2$ doublet were feasible because of their being narrow. However there are no experimental candidates for the $j \! = \! 1/2$ doublet so far; the observation of these two states should be complicated by their large predicted widths.

The experimental results discussed in this review paper concern the recent observation of two states consistent with their identification with the $(0^{+},1^{+})$ missing doublet; they are below $DK$ and $D^{*}K$ thresholds and decay to $D^{+}_{s}\pi^{0}$ and $D^{*+}_{s}\pi^{0}$ final states respectively. 

\section{Observation by BaBar of the narrow state $D^{*}_{sJ}(2317)^{+}$}

\begin{figure}
\includegraphics[height=8.5cm,width=8.7cm]{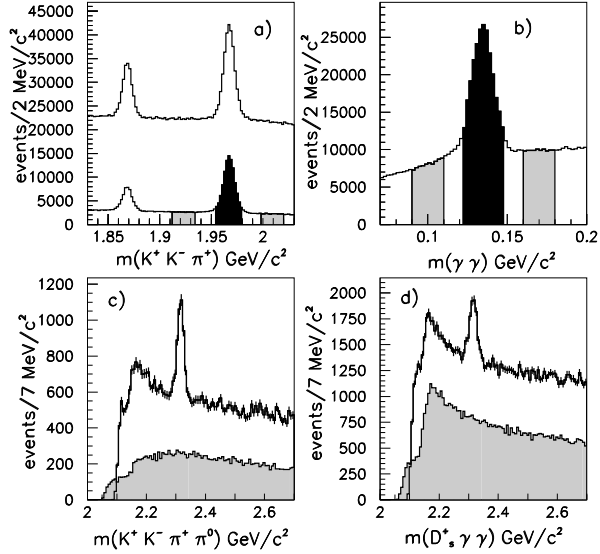}
\caption{The BaBar Collaboration~\cite{babar2317}. \textbf{(a)} The $K^{+}K^{-}\pi^{+}$ mass distribution for all selected candidates; additional selection criteria described in the text have been used to obtain the lower histogram. \textbf{(b)} The mass distribution of two-photon combinations in the selected events. \textbf{(c)} The $K^{+}K^{-}\pi^{+}\pi^{0}$ mass distribution for candidates in the $D^{+}_{s}$ signal/sideband regions of (a) (top/shaded histogram respectively); the $\pi^{0}$ candidates are mass-constrained. \textbf{(d)} The $D^{+}_{s} \gamma \gamma$ mass distribution for candidates in the $D^{+}_{s}$ signal region of (a) coupled with photon pairs from the $\pi^{0}$ signal/sideband regions of (b) (top/shaded histogram respectively); both top and shaded histograms present a broad peak due to random combinations $D^{*+}_{s} \gamma$ where $D^{*+}_{s} \! \rightarrow \! D^{+}_{s} \gamma$.}
\label{fig:fig1}
\end{figure}

The BaBar analysis, described in detail elsewhere~\cite{babar2317}, investigates the inclusively-produced $D^{+}_{s} \pi^{0}$ mass spectrum by combining charged particles corresponding to the decays $D^{+}_{s} \! \rightarrow \! K^{+}K^{-}\pi^{+}$~\cite{chacon} with $\pi^{0}$ candidates that (1) are reconstructed from a pair of photons with energy greater than $100$ MeV and constrained to emanate from the intersection of the $K^{+}K^{-}\pi^{+}$ candidate trajectory and the beam envelope, and (2) are formed by performing a one-constraint fit to the $\pi^{0}$ mass. 
A given event may yield several acceptable $\pi^{0}$ candidates, but only those that do not share any photon with the others are retained. To reduce backgrounds, both that from $B$ meson decays and the combinatorial one from continuum, each $K^{+}K^{-}\pi^{+}\pi^{0}$ candidate is required to have a momentum $p^{*}$ in the $e^{+}e^{-}$ center-of-mass frame greater than $2.5$ \gevc. The upper histogram in Fig.\ref{fig:fig1}(a) shows the $K^{+}K^{-}\pi^{+}$ invariant mass distribution for all candidates characterized by the peaks associated to $D^{+}$ and $D^{+}_{s}$ mesons signal. 

Further background rejection, providing the lower histogram of Fig.\ref{fig:fig1}(a), is achieved by the following additional selection criteria. Quasi-two body decay modes into $\phi \pi^{+}$ and $\overline{K}^{*0}K^{+}$ final states are selected by retaining only those candidates with $K^{+}K^{-}$ mass within $10$ \mevcc of the $\phi(1020)$ mass or with $K^{-}\pi^{+}$ mass within $50$ \mevcc of the $\overline{K}^{*0}(892)$ mass. 
Given the $D^{+}_{s}$ Dalitz plot structure, these candidates belong to two disjoint sub-samples. Moreover the decay products of the vector mesons $\phi(1020)$ and $\overline{K}^{*0}(892)$ exhibit the $cos^{2} \theta_{h}$ behaviour expected because of angular momentum conservation, where $\theta_{h}$ is the helicity angle; the signal-to-background ratio is improved by requiring $|cos \theta_{h}| \! > \! 0.5$.

In Fig.\ref{fig:fig1}(a) the $D^{+}_{s}$ signal and sideband regions are shaded and the $D^{+}_{s}$ signal peak contains about $80,000$ candidates and is centered at a mass of $(1967.20\pm 0.03)$ \mevcc (statistical error only). Fig.\ref{fig:fig1}(b) shows the mass distribution for all two-photon combinations associated with the selected events; the $\pi^{0}$ signal and sideband regions are shaded. Candidates in the $D^{+}_{s}$ signal region of Fig.\ref{fig:fig1}(a) are combined with the mass-constrained $\pi^{0}$ candidates to provide the top histogram of Fig.\ref{fig:fig1}(c). 
The shaded histogram in Fig.\ref{fig:fig1}(c) represents combinations of candidates from the $D^{+}_{s}$ sideband regions with the same $\pi^{0}$ candidates. In Fig.\ref{fig:fig1}(d) the mass distributions result from the combination of candidates from the $D^{+}_{s}$ signal region with the photon pairs from the $\pi^{0}$ signal and sideband regions of Fig.\ref{fig:fig1}(b); the mass resolution is poorer than in Fig.\ref{fig:fig1}(c) because there is no mass constraint for the photon pairs. In order to improve the resolution the nominal $D^{+}_{s}$ mass has been used to calculate the $D^{+}_{s}$ energy.

The top histograms of Fig.\ref{fig:fig1}(c) and Fig.\ref{fig:fig1}(d) (associated to signal-signal combinations) are characterized by a clear narrow signal at a mass close to $2.32$ \gevccc. In the corresponding shaded histograms (associated to signal-sidebands combinations) this signal is absent. These facts indicate unambigously that the peak is associated with the $D^{+}_{s}\pi^{0}$ system. Moreover the signal is observed in both the quasi-two body $D^{+}_{s}$ decay modes. 
Furthermore Monte Carlo simulations have been used to investigate the possibility that the $D^{*}_{sJ}(2317)^{+}$ signal could be due to a reflection from other known charmed states. No hint of any peak is found in the signal region. In addition no peak has been produced by exchanging the kaon and pion identities to simulate particle mis-identification effects.

\begin{figure}
\includegraphics[height=8.6cm,width=8.2cm]{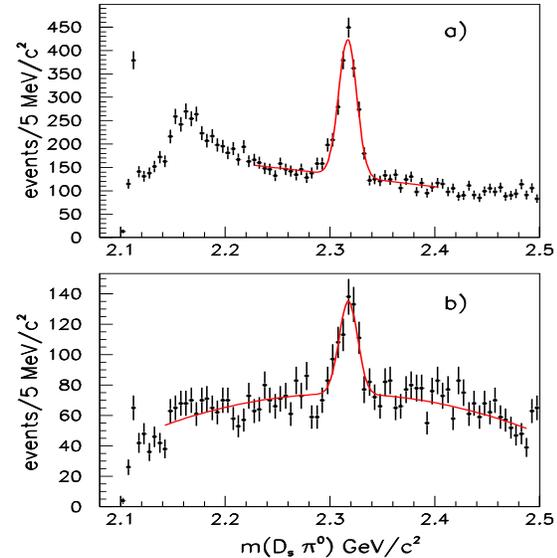}
\caption{The BaBar Collaboration~\cite{babar2317}. The $D^{+}_{s}\pi^{0}$ mass distribution for \textbf{(a)} the decay $D^{+}_{s} \! \rightarrow \! K^{+}K^{-}\pi^{+}$ and \textbf{(b)} the decay $D^{+}_{s} \! \rightarrow \! K^{+}K^{-}\pi^{+}\pi^{0}$. Fit function and results for the $D^{*}_{sJ}(2317)^{+}$ signal are described in the text. The very narrow peak near threshold is associated to the isospin-violating $D^{*+}_{s} \! \rightarrow \! D^{+}_{s} \pi^{0}$ decay.}
\label{fig:fig2}
\end{figure}

Fig.\ref{fig:fig2} shows the $D^{+}_{s}\pi^{0}$ mass distribution for $p^{*}(D^{+}_{s}\pi^{0}) > 3.5$ \gevc for two different $D^{+}_{s}$ decay modes: (a) $D^{+}_{s} \! \rightarrow \! K^{+}K^{-}\pi^{+}$ and (b) $D^{+}_{s} \! \rightarrow \! K^{+}K^{-}\pi^{+}\pi^{0}$. This second additional sample of $D^{+}_{s}$ decays is obtained by adding to each $K^{+}K^{-}\pi^{+}$ combination fitted to a common vertex the $\pi^{0}$ candidates constrained to emanate from this vertex; each resulting $D^{+}_{s}$ candidate is combined with a second $\pi^{0}$ candidate with an harder lab momentum (greater than $300$ \mevccc). Moreover the $K^{*\pm}$, $\overline{K}^{*0}$, $\phi$ or $\rho^{+}$ mass regions are selected for the relevant two-body subsystems. 
The fit function includes a single Gaussian for the signal and a polynomial function for the background distribution. The fit yields $1267\pm 53$ candidates over background for a fitted mass of $2316.8\pm 0.4$ \mevcc with a standard deviation of $8.6\pm 0.4$ \mevcc (statistical errors only), for the $D^{+}_{s}$ decay mode (a). The Gaussian fit for the $D^{+}_{s}$ decay mode (b) is fully consistent with the previous one since it yields $273\pm 33$ candidates with a mean of $2317.6\pm 1.3$ \mevcc and a width of $8.8\pm 1.1$ MeV (statistical errors only). The systematic uncertainty in the mass is conservatively estimated to be less than $3$ \mevcc. 

An estimate of the mass resolution for the $K^{+}K^{-}\pi^{+}\pi^{0}$ system can be provided directly from the data by fitting the mass distribution for $D^{+}_{s} \! \rightarrow \! K^{+}K^{-}\pi^{+}\pi^{0}$ and the measured width is consistent with that of $D^{*}_{sJ}(2317)^{+}$ signal. 
A similar mass resolution can be obtained from the simulation of the $D^{*}_{sJ}(2317)^{+}$ decaying to $K^{+}K^{-}\pi^{+}\pi^{0}$ with a generated intrinsic width close to zero. In other words the width observed in the data is consistent with Monte Carlo expectations for a zero intrinsic width state. Thus the observed width is consistent with the experimental resolution and it can be concluded that the intrinsic width of the $D^{*}_{sJ}(2317)^{+}$ must be below $10$ MeV.

\begin{figure}
\includegraphics[height=7.6cm,width=8.2cm]{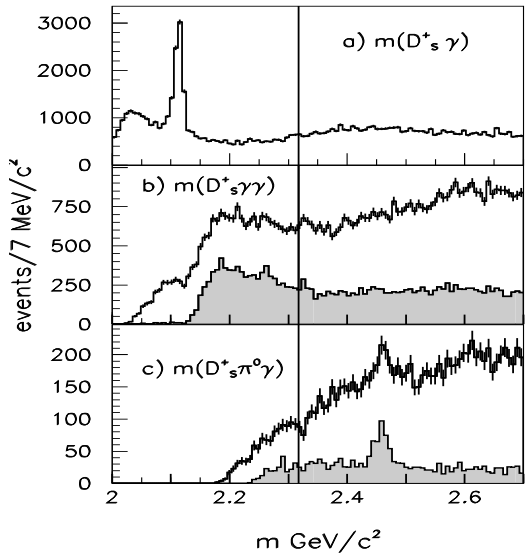}
\caption{The BaBar Collaboration~\cite{babar2317}. The mass distribution for (a) $D^{+}_{s} \gamma$, (b) $D^{+}_{s} \gamma \gamma$ and (c) $D^{+}_{s} \pi^{0} \gamma$. Details on combination criteria are given in the text. The shaded histograms of (b) and (c) correspond to $D^{+}_{s} \gamma$ masses falling in the $D^{*+}_{s}$ signal region. The vertical line indicates the $D^{*}_{sJ}(2317)^{+}$ mass value.}
\label{fig:fig3}
\end{figure}

A search for other $D^{*}_{sJ}(2317)^{+}$ decays has been performed. For this search additional background suppression is achieved by requiring that the $p^{*}$ of each system of particles investigated must be greater than $3.5$ \gevc. Fig.\ref{fig:fig3}(a) shows the $D^{+}_{s} \gamma$ mass distribution obtained by combining a $D^{+}_{s}$ candidate from the signal region of Fig.\ref{fig:fig1}(a) with photons in the selected events not belonging to a $\gamma \gamma$ combination in the signal region of Fig.\ref{fig:fig1}(b) and having an energy greater than $150$ MeV. 
A clear $D^{*+}_{s}$ signal is visible but there is no indication of a $D^{*}_{sJ}(2317)^{+}$ production. No signal near $2.32$ \gevcc is observed in the $D^{+}_{s} \gamma \gamma$ mass distribution presented in Fig.\ref{fig:fig3}(b) as well; again any photon belonging to a $\gamma \gamma$ combination in the $\pi^{0}$ signal region is excluded here. The shaded histogram of Fig.\ref{fig:fig3}(b) is associated to the subset of combinations for which either $D^{+}_{s} \gamma$ combination falls in the $D^{*+}_{s}$ signal region (defined as $2.096 \! < \! m(D^{+}_{s}\gamma) \! < \! 2.128$ \gevccc). A $D^{*}_{sJ}(2317)^{+}$ signal is missing here too; this implies the absence of a $D^{*+}_{s} \gamma$ decay mode at the present level of statistics. 

\begin{figure}
\includegraphics[height=6cm,width=9.1cm]{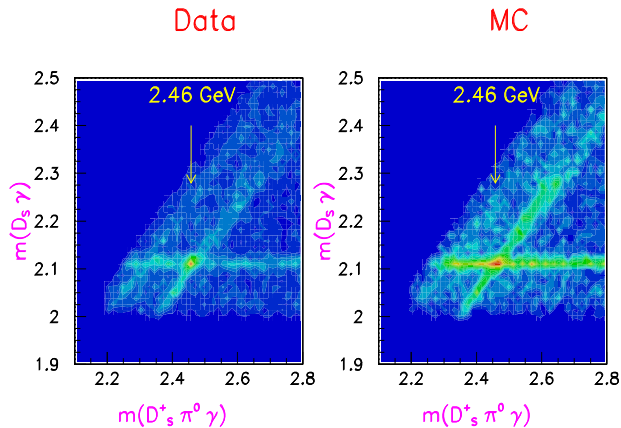}
\caption{The BaBar Collaboration. \textbf{(Right)} Simulated data $e^{+}e^{-} \! \rightarrow \! c\overline{c}$ including $D^{*}_{sJ}(2317)^{+}$. The scatter plot shows how a peaking background, at about $2.46$ \gevcc in the $D^{+}_{s} \pi^{0} \gamma$ mass projection, can be generated by the kinematical overlap of the $D^{*+}_{s} \! \rightarrow \! D^{+}_{s} \gamma$ signal band with the $D^{*}_{sJ}(2317)^{+} \! \rightarrow \! D^{+}_{s} \pi^{0}$ signal band obtained adding respectively an unrelated $\pi^{0}$ and an unrelated $\gamma$. \textbf{(Left)} Real data are characterized, at least partially, by the same kinematical cross-through.}
\label{fig:fig4}
\end{figure}

The top histogram of Fig.\ref{fig:fig3}(c) shows the $D^{+}_{s} \pi^{0} \gamma$ mass distribution excluding again any photon that belongs to the $\pi^{0}$ signal region. The subset of combinations in which the $D^{+}_{s} \gamma$ mass falls in the $D^{*+}_{s}$ signal region is represented by the shaded histogram of Fig.\ref{fig:fig3}(c). No signal near $2.32$ \gevcc is observed in both distributions. However a peak is visible near a mass value of $2.46$ \gevcc and survives at the $D^{*+}_{s}$ selection. 
This mass corresponds to the overlap region of the $D^{*+}_{s} \! \rightarrow \! D^{+}_{s} \gamma$ and $D^{*}_{sJ}(2317)^{+} \! \rightarrow \! D^{+}_{s} \pi^{0}$ signal bands that produces, because of the small widths of both the $D^{*+}_{s}$ and $D^{*}_{sJ}(2317)^{+}$ mesons, a narrow peak in the $D^{+}_{s} \pi^{0} \gamma$ mass distribution. It has been checked the possibility that this peak at about $2.46$ \gevcc would kinematically feed-down the $D^{*}_{sJ}(2317)^{+}$ signal, provided that it could eventually be a new additional narrow state decaying to $D^{*+}_{s} \pi^{0}$. 
This eventual feed-down would be produced as follows: $e^{+}e^{-} \! \rightarrow \! X(2460)^{+} X_{recoil}$, $X(2460)^{+} \! \rightarrow \! D^{*+}_{s} \pi^{0}$ and $D^{*+}_{s} \! \rightarrow \! D^{+}_{s} \gamma$ where the $D^{+}_{s} \pi^{0}$ combination could emulate a $D^{*}_{sJ}(2317)^{+}$ candidate and the residual photon would appear to be unrelated or simply lost. Such $D^{+}_{s} \pi^{0}$ reflected peak, however, would have a root-mean-square of about $15$ \mevcc, namely quite twice the width of the observed $D^{*}_{sJ}(2317)^{+}$ signal. The second crucial check is obtained by a Monte Carlo unfolding method: if the apparent signal at $2.46$ \gevcc were due to a state decaying entirely to $D^{*+}_{s} \pi^{0}$, it would produce only one-sixth of the observed signal at $2.32$ \gevcc.

\section{Confirmation of $D^{*}_{sJ}(2317)^{+}$ and observation of $D^{*}_{sJ}(2458)^{+}$ by CLEO and Belle}

\begin{figure}
\includegraphics[width=\hsize]{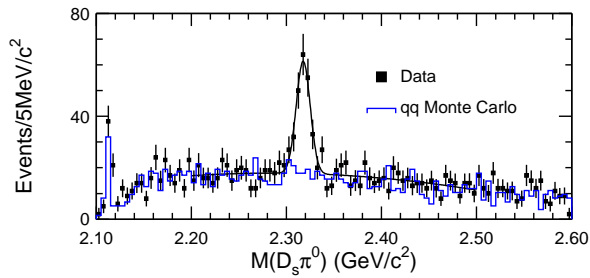}
\caption{The CLEO Collaboration~\cite{cleo2317}. $\Delta m(D^{+}_{s}\pi^{0})$ mass difference distribution. The points represent the data; the solid histogram is the predicted normalized spectrum from simulated $e^{+}e^{-} \! \rightarrow \! q\overline{q}$ events.}
\label{fig:fig5}
\end{figure}

\begin{figure}
\includegraphics[height=5.2cm,width=6.4cm]{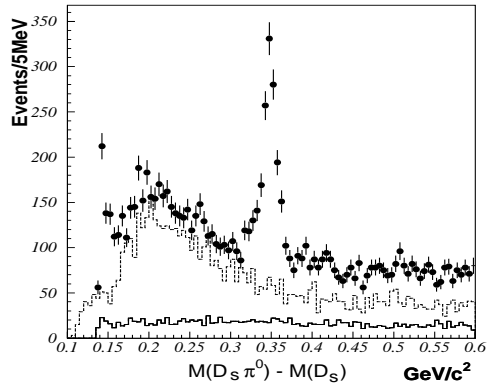}
\caption{The Belle Collaboration~\cite{belle2317}. $\Delta m(D^{+}_{s}\pi^{0})$ mass difference distribution. The points represent the data; the histograms are associated to $D^{+}_{s}$ (solid) and $\pi^{0}$ (dashed) sideband regions.}
\label{fig:fig6}
\end{figure}

CLEO~\cite{cleo2317} and Belle~\cite{belle2317} have readily confirmed the $D^{*}_{sJ}(2317)^{+}$ state seen by BaBar, using only the $\phi \pi^{+}$ quasi two-body $D^{+}_{s}$ decay mode. In terms of the mass difference $\Delta m (D^{+}_{s}\pi^{0}) \! = \! m(K^{+}K^{-}\pi^{+}\pi^{0})-m(K^{+}K^{-}\pi^{+})$, CLEO reports a value of $\Delta m = 350.3\pm 1.0$ \mevcc with $165\pm 20$ candidates (Fig.\ref{fig:fig5}), whereas Belle reports $\Delta m = 348.7\pm 0.5$ \mevcc with $761\pm 44$ candidates (Fig.\ref{fig:fig6}). These values (with statistical errors only) are in good agreement with the BaBar measurement of $\Delta m = 348.4\pm 0.4$ \mevcc ~\cite{antimo2458}. Using their Monte Carlo simulation CLEO is able to limit the natural width of the $D^{*}_{sJ}(2317)^{+}$ state to be less than $7$ MeV at the $90\%$ confidence level (C.L.).

As just discussed, in order to establish the existence of the $D^{*}_{sJ}(2317)^{+}$, BaBar has ruled out any relevant feeding down by a state of mass $2.46$ \gevcc eventually produced in addition to the $D^{*}_{sJ}(2317)^{+}$. On the other hand BaBar has been rather careful to state that the structure observed in the $D^{+}_{s} \pi^{0} \gamma$ mass spectrum is another new narrow state; the complexity of the overlapping kinematics of the $D^{*+}_{s} \! \rightarrow \! D^{+}_{s} \gamma$ and $D^{*}_{sJ}(2317)^{+} \! \rightarrow \! D^{+}_{s} \pi^{0}$ decays has suggested a dedicated and detailed study to evaluate the amount of feeding-up from the $D^{*}_{sJ}(2317)^{+}$. 
Fig.\ref{fig:fig4}(right) illustrates, by means of a Monte Carlo study, the kinematical cross-through: the $D^{*+}_{s} \! \rightarrow \! D^{+}_{s} \gamma$ signal band, obtained by adding a random $\pi^{0}$, crosses the band produced by $D^{*}_{sJ}(2317)^{+} \! \rightarrow \! D^{+}_{s} \pi^{0}$ plus an additional random $\gamma$, thus generating a peak at about $2.46$ \gevcc in the $D^{+}_{s} \pi^{0} \gamma$ mass projection. This pattern also characterizes, at least partially, the same scatter plot for the real data in Fig.\ref{fig:fig4}(left).

\begin{figure}[t]
\includegraphics[height=6cm,width=7cm]{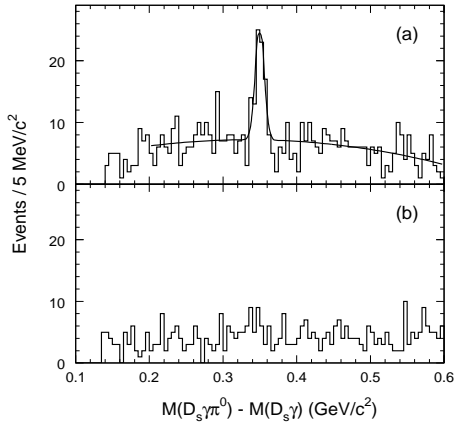}
\caption{The CLEO Collaboration~\cite{cleo2317}. $\Delta m (D^{*+}_{s}\pi^{0})$ mass difference distributions, for \textbf{(a)} $D^{*+}_{s}$ signal region and \textbf{(b)} $D^{*+}_{s}$ sideband regions. The signal yield obtained by the fit is $55\pm 10$ candidates.}
\label{fig:fig7}
\end{figure}

\begin{figure}
\includegraphics[height=5.5cm,width=6.5cm]{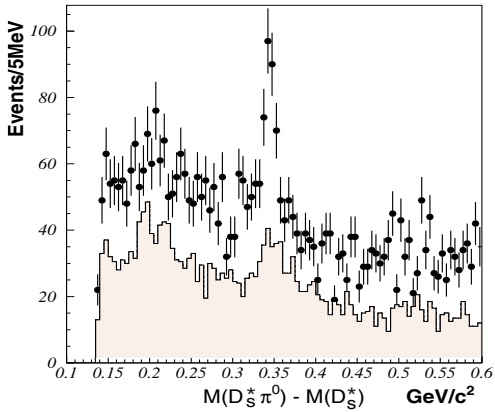}
\caption{The Belle Collaboration~\cite{belle2317}. $\Delta m (D^{*+}_{s}\pi^{0})$ mass difference distributions showing the evidence for $D^{*}_{sJ}(2458)$ state. The shaded histogram is associated to the $D^{*+}_{s}$ sidebands. The signal yield obtained by a fit (not shown here) is $126\pm 25$ candidates.}
\label{fig:fig8}
\end{figure}

CLEO~\cite{cleo2317} and Belle~\cite{belle2317} have analyzed the $D^{*+}_{s} \pi^{0}$ mass spectrum finding evidence for the structure in the $2.46$ \gevcc region and claiming it is a new state, the $D^{*}_{sJ}(2458)^{+}$. The mass difference is now defined as $\Delta m (D^{*+}_{s}\pi^{0}) = m(K^{+}K^{-}\pi^{+}\pi^{0} \gamma)-m(K^{+}K^{-}\pi^{+}\gamma)$ and its distribution is provided by CLEO in Fig.\ref{fig:fig7} and by Belle in Fig.\ref{fig:fig8}. CLEO has addressed the problem of feed-up in two different ways. By an unfolding method using simulated data they estimated a rather small amount of feed-up (about $9\%$) which is consistent with the small peak in the $D^{*+}_{s}$ sidebands (Fig.\ref{fig:fig7}(b)). On the other hand CLEO fits the signal peak after performing a $D^{*+}_{s}$ sidebands subtraction. 
The two methods provide consistent results in terms of $D^{*}_{sJ}(2458)^{+}$ candidates yield and the mean and the width values of the fitted $D^{*}_{sJ}(2458)^{+}$ signal peak in $\Delta m (D^{*+}_{s}\pi^{0})$ distribution. Belle's observation~\cite{belle2317} of the $D^{*}_{sJ}(2458)^{+}$ state is given in Fig.\ref{fig:fig8} where there is also visible a clear peaking in the $D^{*+}_{s}$ sideband regions that shows a relevant level of feed-up (about $30\%$~\cite{bellecipanp}). Afterwards BaBar has confirmed the existence of the $D^{*}_{sJ}(2458)^{+}$ state~\cite{antimo2458}~\cite{babar2458}. The $D^{*}_{sJ}(2458)^{+}$ mass measurements reported by Belle~\cite{belle2317} and BaBar~\cite{babar2458} are in agreement: $2456.5\pm 1.7$ \mevcc and $2458.0\pm 1.4$ \mevcc respectively (with combined statistical and systematic uncertainties); however the mass value obtained by CLEO is two standard deviations larger: $2463.5\pm 2.1$ \mevcc. All three experiments have used Monte Carlo simulations to evaluate the detector mass resolution. CLEO~\cite{cleo2317} provides a $90\%$ C.L. upper limit of $7$ MeV on the natural width of the $D^{*}_{sJ}(2458)^{+}$ state.  

\begin{figure}
\includegraphics[height=6.5cm,width=7.2cm]{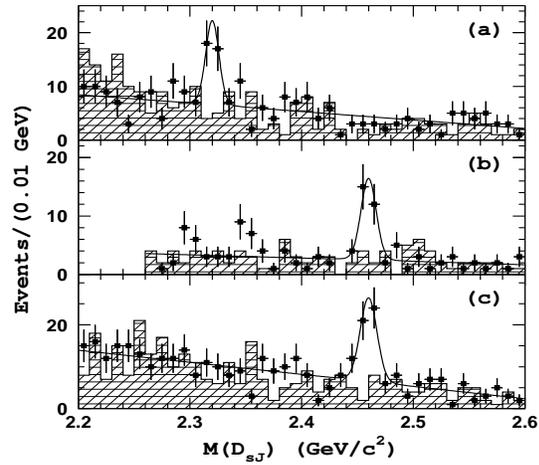}
\caption{The Belle Collaboration~\cite{bellebdecays}. Invariant mass distributions for $D^{*}_{sJ}$ candidates produced in the reaction $B \! \rightarrow \! \overline{D} D^{*}_{sJ}$ and reconstructed in the three decay modes \textbf{(a)} $D_{s} \pi^{0}$, \textbf{(b)} $D^{*}_{s} \pi^{0}$ and \textbf{(c)} $D_{s} \gamma$. Points with errors represent the candidates associated to the $\Delta E$ signal region whereas hatched histograms those falling into $\Delta E$ sidebands; $\Delta E = (\sum_{i} E_{i})- E_{beam}$ where $E_{beam}=\sqrt{s}/2$ is the beam energy and $E_{i}$ are the energies of the decay products of the $B$ meson in the center-of-mass frame.}
\label{fig:fig9}
\end{figure}

These two new states have been observed for the first time by Belle also in B decays~\cite{bellebdecays}. Fig.\ref{fig:fig9} shows the reconstructed $D^{+}_{s} \pi^{0}$, $D^{*+}_{s} \pi^{0}$ and $D^{+}_{s} \gamma$ mass spectra for candidates whose mass and beam energy constraints are consistent with the decay $B \! \rightarrow \! \overline{D} D^{*}_{sJ}$. The signal in Fig.\ref{fig:fig9}(c) represents the first observation of the radiative decay $D^{*}_{sJ}(2458)^{+} \! \rightarrow \! D^{+}_{s} \gamma$. From the two $B \! \rightarrow \! \overline{D} D^{*}_{sJ}(2458)$ branching fraction measurements, Belle determines the ratio

\noindent
\begin{center}
%\begin{equation}
\[
\frac{\mathcal{B} (D^{*}_{sJ}(2458) \! \rightarrow \! D_{s} \gamma)}{\mathcal{B} (D^{*}_{sJ}(2458) \! \rightarrow \! D^{*}_{s} \pi^{0})} = 0.38 \pm 0.11 \mbox{(stat.)} \pm 0.04 \mbox{(syst.)}
\label{eq:radratio}
\]
%\end{equation}
\end{center}

The Belle measurement of this ratio for the $c\overline{c}$ events~\cite{belle2317} is consistent with the previous one: $0.55 \pm 0.13 \mbox{(stat.)} \pm 0.08 \mbox{(syst.)}$.

For what concerns the $D^{*}_{sJ}(2458)^{+}$ mass determination from $B$ decays, Belle obtaines a value of $2459.2 \pm 2.6$ \mevcc (with combined statistical and systematic uncertainties) which is compatible with the measurements from continuum events either by CLEO and by BaBar and themselves. Using $B$ reconstruction to obtain mass values for both states may be the cleanest method since the feed across corrections are eluded.

\section{Properties of the new $D_{sJ}$ states and discussion about their nature.}

There is no question about the existence of such two new states but their nature need still to be investigated. 

The masses of the new states appear surprisingly low for quark model spectroscopy; specifically $D^{*}_{sJ}(2317)^{+}$ is below $DK$ threshold and $D^{*}_{sJ}(2458)^{+}$ falls below $D^{*}K$ one. If they are interpreted as ordinary $c \overline{s}$ states, this fact forces them to decay mainly via isospin-violating transition making their widths quite narrow. Isospin is violated in $D^{*}_{sJ}(2317)^{+} \! \rightarrow \! D^{+}_{s}\pi^{0}$ and $D^{*}_{sJ}(2458)^{+} \! \rightarrow \! D^{*+}_{s}\pi^{0}$ since the $D^{+}_{s}$ and its excitations are $I \! = \! 0$, while the pion is $I \! = \! 1$. A possible mechanism is the decay through a virtual $\eta$ followed by $\eta-\pi^{0}$ mixing ~\cite{chowise}.

After the announcement of the $D^{*}_{sJ}(2317)^{+}$ discovery several possible interpretations have appeared: since the $D^{*}_{sJ}(2317)^{+}$ contradicts the current models of charmed mesons spectroscopy either these models need modification or the observed state might have a non-$q\overline{q}$ nature. Among the ``exotic'' explanations, a $DK$ molecule~\cite{barnes}, a four-quark state~\cite{4quark} and a $D\pi$ atom~\cite{szc} have been proposed. Among ``ordinary'' explanations there have been attempts to revise~\cite{godfrey} or modify~\cite{cahn} potential models and to couple chiral perturbation theory with HQET~\cite{bardeen}~\cite{nowak}~\cite{coldefa}~\cite{polosa}. Recently it has been even proposed that these states are a mixture of $c\overline{s}$ and four-quark states~\cite{pakpetrov}.

On experimental side, in order to clarify the nature of these states it is necessary to determine their quantum numbers and decay branching fractions, particularly those for radiative decays. The spin-parity of these states can be inferred from their decay modes. 

For a parity-conserving decay of $D^{*}_{sJ}(2317)^{+}$ to $D^{+}_{s}\pi^{0}$ only a natural spin-parity series is allowed, that is $J^{P} \! = \! \{0^{+},1^{-},2^{+},...\}$. $J^{P} \! = \! 0^{+}$ is suggested by considering at the same time that (1) the mass is low compared to that of the $D_{s1}(2536)^{+}$ and $D^{*}_{sJ}(2573)^{+}$ and (2) the efficiency corrected helicity distribution is consistent with being flat~\cite{babar2317}, as expected for a spin-zero particle or for a particle of higher spin that is produced unpolarized. If $J^{P} \! = \! 0^{+}$ the $D^{*}_{sJ}(2317)^{+}$ cannot decay to $D^{+}_{s}\gamma$, whereas the decay to $D^{*+}_{s}\gamma$ is allowed by parity and angular momentum conservations. Moreover it could not decay into three pseudoscalars. The absence of the $D^{+}_{s}\gamma$~\cite{babar2317}~\cite{cleo2317}~\cite{belle2317} and of the $D^{+}_{s} \pi^{+} \pi^{-}$~\cite{cleo2317}~\cite{belle2317} decay modes makes a $J^{P} \! = \! 0^{+}$ assignment most likely. 
%%%The apparent absence of the decay to $D^{*+}_{s}\gamma$~\cite{babar2317}~\cite{cleo2317}~\cite{belle2317} may simply indicate that decay by pion emission is favoured over radiative decay. 
CLEO~\cite{stoneurheim} has also looked for neutrally charged states in $D^{\pm}_{s}\pi^{\mp}$ and doubly charged states in $D^{\pm}_{s}\pi^{\pm}$ without finding any signal. This is an argument against any molecular interpretation.

On the other hand for the $D^{*}_{sJ}(2458)^{+}$ an unnatural spin-parity is suggested and  thus this state is consistent with being the missing $J^{P} \! = \! 1^{+}$ P-wave state. The observation of the $D^{*}_{sJ}(2458)^{+} \! \rightarrow \! D^{+}_{s}\gamma$ decay~\cite{belle2317}~\cite{bellebdecays} rules out the $J \! = \! 0$ assignment and favours a $1^{+}$ interpretation. Angular distributions from the $B$ decays to this state indicate directly that spin one is strongly preferred~\cite{bellebdecays}. Moreover $J \! = \! 1^{+}$ allows a strong, isospin-conserving but OZI-suppressed, decay to $D^{+}_{s} \pi^{+} \pi^{-}$. Indeed this di-pion decay mode has been observed by Belle~\cite{belle2317}. The apparent absence of the decay $D^{*}_{sJ}(2458)^{+} \! \rightarrow \! D^{*}_{sJ}(2317)^{+} \gamma$~\cite{cleo2317}~\cite{babar2458} may indicate that the electromagnetic decay mechanism cannot compete with a likely strong but isospin-violating process resulting from $\eta-\pi^{0}$ mixing.

In summary it must be pointed out that most of the properties of the new states, $D^{*}_{sJ}(2317)^{+}$ and $D^{*}_{sJ}(2458)^{+}$, can be explained if these particles are $c\overline{s}$ states. The most likely assignment for their spin-parity is $0^{+}$ and $1^{+}$. Models based on chiral simmetry coupled to HQET~\cite{bardeen}~\cite{nowak} predict that these $0^{+}$ and $1^{+}$ are the chiral partners of the $0^{-}$ and $1^{-}$ states. This ``parity doubling'' prediction is experimentally confirmed: the chiral mass splitting between the $0^{+}$ and $0^{-}$ states, that is $m(D^{*}_{sJ}(2317)^{+})-m(D^{+}_{s})$, is equal within experimental error to that between the $1^{+}$ and $1^{-}$ states, namely $m(D^{*}_{sJ}(2458)^{+})-m(D^{*+}_{s})$. Moreover the expected rates for radiative decays in the same models are mostly consistent with experimental measurements or limits. 

Further experimental information coupled with theoretical ideas can definitively shed light on the nature of these new mesons.

\section*{Acknowledgments}

Many thanks to A.~Palano for his helpful suggestions.


\begin{thebibliography}{9}
\bibitem{chacon} The inclusion of charge-conjugate configurations is implied throughout this paper.
\end{thebibliography}
\end{document}